\documentclass[11pt,aps]{article}
\usepackage{amsfonts,amssymb,eucal}
\usepackage{amsthm}
 \usepackage{amsmath}
\usepackage{amscd}
 \usepackage{latexsym} 

\newcommand{\R}{\ensuremath{\mathbb{R}}}
\newcommand{\C}{\ensuremath{\mathbb{C}}}

\begin{document}

\title{\bf Circulant conference matrices for new complex  Hadamard matrices}
\author{Petre Di\c t\u a\\
National Institute of Physics and Nuclear Engineering,\\
M\u agurele, Ilfov, P.O. Box MG6, Romania\\ email: dita@zeus.theory.nipne.ro}

\maketitle

\begin{abstract}
The circulant real and complex matrices are  used to find new real and complex conference matrices. With them we construct Sylvester inverse orthogonal matrices  by doubling the size of inverse complex conference matrices.   When the free  parameters  take values on the unit circle the inverse orthogonal matrices transform into complex Hadamard matrices. The method is used for $n=6$ conference matrices and in this way we find new parametrisations of Hadamard matrices for dimension $ n=12$.
\end{abstract}

\section{Introduction}

The inverse orthogonal matrices in the most general form have been defined  by   Sylvester  in  paper \cite{JJS}. In our   paper   we shall make use only   of a   particular class of  inverse orthogonal matrices, $A=(a_{ij})$, namely   those matrices whose inverse  is given by 
$A^{-1}=(1/a_{ij})^t=(1/a_{ji})$, where $t$ means transpose, and their entries $0\ne a_{ij}\in \C$  satisfy the relation
\begin{eqnarray}
A A^{-1}= n I_n \label{inv}
\end{eqnarray}
In the above relation $I_n$ is the $n$-dimensional identity matrix. When the entries  $a_{ij}$ take values on the unit circle, $A^{-1}$ coincides with the Hermitian conjugate $A^*$ of $A$, and in this case (\ref{inv}) is the definition of  complex Hadamard matrices. Complex Hadamard matrices have applications in quantum information theory, several branches of combinatorics, engineering, statistics, digital signal processing, image analysis, coding theory, cryptology, etc. 

Complex orthogonal matrices  appeared in the description of topological invariants of knots and links, see e.g. \cite{VFRJ}, which under the name of two-weight spin models have been  related to symmetric statistical Potts models. By removing the symmetry condition they led  to two-weight spin models, \cite{KMW}, and a new generalization provided  four-weight spin models, \cite{BB}. These last  matrices are also known under the name of  type II matrices, \cite{KN}.
Particular cases of type II matrices also appeared as  generalized Hadamard transform for processing  multiphase or   multilevel signals, see \cite{LRP} and \cite{H}, which includes the classical Fourier, Walsh-Hadamard and Reverse Jacket transforms. They are defined for $2n\times 2n$-dimensional Butson type matrices,
\cite{B}, i.e. in terms of   $p^{th}$ roots of unity and/or in one or more  complex non-zero parameters \cite{LRP}, \cite{BAS}. 

In the paper we provide an analytic method for the construction of complex  conference matrices that lead to inverse orthogonal matrices, and, as a byproduct,  to  nonequivalent complex Hadamard matrices.     The  method we use in the following to construct complex Hadamard matrices is via doubling the size of inverse complex conference matrices,  and  here we treat the case   $n=6$  to obtain new complex Hadamard matrices of dimension $n=12$.

\section{Circulant Matrices}

A $n\times n$ circulant matrix is a matrix whose every row is a right cyclic shift of the preceding one so that  its form can be written as 
\begin{eqnarray}
\mathfrak{C}_n=\left[ \begin{array}{ccccc}
c_0&c_1&c_2& \ldots & c_{n-1}\\
c_{n-1}&c_0&c_1&\ldots &c_{n-2}\\
c_{n-2}&c_{n-1}&c_0&&\vdots\\
\vdots&&&\ddots&\\
c_1&c_2&&\ldots&c_0\end{array}\right]\label{circ}
\end{eqnarray}

For getting conference matrices, denoted in the following by $C_n$,  we take $c_0=0$ and we have to find those constraints on parameters $c_i,\,\,i=1,\ldots,n-1$,  for which (\ref{circ}) gets a conference matrix.

Real conference matrices were firstly constructed by Paley, \cite{P}, although their names come from their application to conference telephony, \cite{Be}. A different approach to find $C_n$ was provided by Goethals and Seidel, \cite{GE}, whose central problem was the construction of symmetric and skew-symmetric real conference matrices for arbitrary $n$.

 The complex  $n\times n$
 conference matrices, $ C_n$,  are the matrices with $a_{ii}=0,\,\, i=1,\dots,n$ and
 $|a_{ij}|=1, \,i\ne j$ that  satisfy
\begin{eqnarray}
C_n\,C_n^* =(n-1)I_n\label{con}\end{eqnarray}
where $C_n^*$ is the Hermitian conjugate of $C_n$.
Complex conference matrices, $C_n$,  are important because by  construction the matrix
\begin{eqnarray}
H_{2n} = \left[\begin{array}{cr}
C_n +I_n&C_n^* -I_n\\
C_n -I_n&-C_n^* -I_n\end{array}\right]\label{conf}
\end{eqnarray}
  is  complex Hadamard of size $ 2n\times 2n$, as one can easily verify.

In this paper we discuss only the case $n=6$ that leads to the construction of twelve-dimensional  complex Hadamard matrices. 
 Usually these matrices have the  entries of the first row and  column  equal to $1$, with one exception, their first entry being $0$.

 Supposing we have a complex conference matrix, $C_n$, that depends   on a few arbitrary phases $e^{i \alpha_j},\,\,j=1,...,k$,  it  can be transformed into a complex inverse orthogonal conference  matrix by the change $e^{i \alpha_j}\rightarrow a_j $ with $0\ne a_j\in$\,{\C} complex non-zero numbers. Thus the  complex inverse  orthogonal conference matrices are  those matrices   with $a_{ij},\,i\ne j$, complex non-zero numbers, and $a_{ii}=0$, defined  by a similar relation to the relation (\ref{con}). It is well known that for complex Hadamard and conference matrices the Hermitian conjugate coincides with the inverse matrix. We extend the  above construction (\ref{conf}) to complex inverse orthogonal conference matrices by providing a formula for $C_n^{-1} $ which has the following form
\begin{eqnarray}
C_n^{-1}=(1/(C_n + I_n)-I_n)^t\label{W1}\end{eqnarray}
where  $1/A$ is the matrix whose elements are $1/a_{ij}$.
In the case of complex orthogonal matrices the formula (\ref{conf}) takes the form
\begin{eqnarray}
O_{2n} = \left[\begin{array}{cr}
C_n +I_n&C_n^{-1} -I_n\\
C_n -I_n&-C_n^{-1} -I_n\end{array}\right]\label{ort}
\end{eqnarray}

For $n=6$   there are a few different  complex orthogonal conference matrices, but most of them lead to equivalent Hadamard matrices. However there a few ones that could be transformed into  nonequivalent Hadamard matrices. From our paper  \cite{D}  we make  use of the $C_6$ matrix which depends on two arbitrary phases that lead to the following complex orthogonal conference matrix.
\begin{eqnarray}
C_{6}=
\left[\begin{array}{crrrrr}
0&1&1&1&1&1\\
1&0&p &p&-q &-q \\
1&p &0&-p&p/q&-p/q\\
1&p &-p&0&-p/q &p/q \\
1&-p&p\, q&-p\,q &0&p\\
1&-p &-p\,q &p\,q &p&0\\
\end{array}
\right]\label{conf6}
\end{eqnarray}
where $p$ and $q$ are complex non zero numbers. In fact one can 
take them from the set $p = \pm 1,\,q =\pm 1$. When $p=q=1$ one gets the conference matrix
\begin{eqnarray}
C_{6a}=
\left[\begin{array}{crrrrr}
0&1&1&1&1&1\\
1&0&1 &1&-1 &-1 \\
1&1 &0&-1&1&-1\\
1&1 &-1&0&-1 &1 \\
1&-1&1 &-1 &0&1\\
1&-1 &-1 &1 &1&0\\
\end{array}
\right]\label{conf6a}
\end{eqnarray}
The choice  $p=1, q=-1$ leads to
\begin{eqnarray}
C_{6b}=
\left[\begin{array}{crrrrr}
0&1&1&1&1&1\\
1&0&1 &1&-1 &-1 \\
1&1 &0&-1&-1&1\\
1&1 &-1&0&1 &-1 \\
1&-1&-1 &1 &0&1\\
1&-1 &1 &-1 &1&0\\
\end{array}
\right]\label{conf6b}
\end{eqnarray}
It is easily seen that $C_{6a}$ and $C_{6b}$ are not equivalent, but more important is that   both of them  lead to nonequivalent complex Hadamard matrices.

The corresponding Paley matrix, \cite{P}, for $n=6$ has the form
\begin{eqnarray}
C_{6c}=
\left[\begin{array}{crrrrr}
0&1&1&1&1&1\\
1&0&1 &-1&-1& 1 \\
1&1 &0&1&-1&-1\\
1&-1 &1&0&1 &-1 \\
1&-1&-1 &1 &0&1\\
1&1 &-1 &-1 &1&0\\
\end{array}
\right]\label{conf6c}
\end{eqnarray}

All the above three matrices are symmetric.
It is easily seen that   the second column of the matrix (\ref{ort}) acts as a symmetry breaking in the doubling formula (\ref{ort}), and by consequence we can multiply the columns, or the rows, of $C_n$  by arbitrary non-null complex numbers. Thus in  the above case and those similar to  we multiply at right the matrices (\ref{conf6a})-(\ref{conf6c})  by the diagonal matrix generated by $(a_1,a_2,a_3,a_4,a_5,a_6)$, where $a_i$  are non-zero complex numbers. In general, by taking into account the nonlinear dependence of the formula (\ref{ort}) on $C_n$ entries  we will   obtain   complex orthogonal matrices  depending on $n$  free parameters, where $n$ denotes the corresponding dimension of the matrix. In our case,  $n=6$, the expectation is to find   orthogonal matrices that depend on six complex non zero numbers.

The above three matrices, (\ref{conf6a}), (\ref{conf6b})  and (\ref{conf6c}), lead to the inverse orthogonal matrices (\ref{O12a}), (\ref{O12b}) and (\ref{O12c}) which depend on six complex non-zero parameters, and to the complex Hadamard matrices of the form (\ref{h12a}), (\ref{h12b})  and  (\ref{h12c}) when the parameters ($a,b,c,d,e,f$) $\in$ \R$^6$, where

\begin{eqnarray}
 O_{12a}(a,b,c,d,e,f)=
\small{\left[\begin{array}{crrrrrrrrrrr}
1&1&1&1&1&1&1&1&1&1&1&1\\*[2mm]
1&b&a&a&-a&-a&-b&-1&\frac{b}{a}&\frac{b}{a}&-\frac{b}{a}&-\frac{b}{a}\\*[2mm]
1&a&c&-a&a&-a&-c&\frac{c}{a}&-1&-\frac{c}{a}&\frac{c}{a}&-\frac{c}{a}\\*[2mm]
1&a&-a&d&-a&a&-d&\frac{d}{a}&-\frac{d}{a}&-1&-\frac{d}{a}&\frac{d}{a}\\*[2mm]
1&-a&a&-a&e&a&-e&-\frac{e}{a}&\frac{e}{a}&-\frac{e}{a}&-1&\frac{e}{a}\\*[2mm]
1&-a&-a&a&a&f&-f&-\frac{f}{a}&-\frac{f}{a}&\frac{f}{a} &\frac{f}{a}&-1\\*[2mm]
1&-b&a&a&-a&-a&b&-1&-\frac{b}{a}&-\frac{b}{a}&\frac{b}{a}&\frac{b}{a}\\*[2mm]
1&a&-c&-a&a&-a&c&-\frac{c}{a}&-1&\frac{c}{a}&-\frac{c}{a}&\frac{c}{a}\\*[2mm]
1&a&-a&-d&-a&a&d&-\frac{d}{a}&\frac{d}{a}&-1&\frac{d}{a}&-\frac{d}{a}\\*[2mm]
1&-a&a&-a&-e&a&e&\frac{e}{a}&-\frac{e}{a}&\frac{e}{a}&-1&-\frac{e}{a}\\*[2mm]
1&-a&-a&a&a&-f&f&\frac{f}{a}&\frac{f}{a}&-\frac{f}{a} &-\frac{f}{a}&-1\\*[2mm]
1&-1&-1&-1&-1&-1&-1&1&1&1&1&1
\end{array}
\right]}\label{O12a}
\end{eqnarray}

\begin{eqnarray}
 O_{12b}(a,b,c,d,e,f)=
\small{\left[\begin{array}{crrrrrrrrrrr}
1&1&1&1&1&1&1&1&1&1&1&1\\*[2mm]
1&b&a&a&-a&-a&-b&-1&\frac{b}{a}&\frac{b}{a}&-\frac{b}{a}&-\frac{b}{a}\\*[2mm]
1&a&c&-a&-a&a&-c&\frac{c}{a}&-1&-\frac{c}{a}&-\frac{c}{a}&\frac{c}{a}\\*[2mm]
1&a&-a&d&a&-a&-d&\frac{d}{a}&-\frac{d}{a}&-1&\frac{d}{a}&-\frac{d}{a}\\*[2mm]
1&-a&-a&a&e&a&-e&-\frac{e}{a}&-\frac{e}{a}&\frac{e}{a}&-1&\frac{e}{a}\\*[2mm]
1&-a&a&-a&a&f&-f&-\frac{f}{a}&\frac{f}{a}&-\frac{f}{a} &\frac{f}{a}&-1\\*[2mm]
1&-b&a&a&-a&-a&b&-1&-\frac{b}{a}&-\frac{b}{a}&\frac{b}{a}&\frac{b}{a}\\*[2mm]
1&a&-c&-a&-a&a&c&-\frac{c}{a}&-1&\frac{c}{a}&\frac{c}{a}&-\frac{c}{a}\\*[2mm]
1&a&-a&-d&a&-a&d&-\frac{d}{a}&\frac{d}{a}&-1&-\frac{d}{a}&\frac{d}{a}\\*[2mm]
1&-a&-a&a&-e&a&e&\frac{e}{a}&\frac{e}{a}&-\frac{e}{a}&-1&-\frac{e}{a}\\*[2mm]
1&-a&a&-a&a&-f&f&\frac{f}{a}&-\frac{f}{a}&\frac{f}{a} &-\frac{f}{a}&-1\\*[2mm]
1&-1&-1&-1&-1&-1&-1&1&1&1&1&1
\end{array}
\right]}\label{O12b}
\end{eqnarray}

\begin{eqnarray}
 O_{12c}(a,b,c,d,e,f)=
\small{\left[\begin{array}{crrrrrrrrrrr}
1&1&1&1&1&1&1&1&1&1&1&1\\*[2mm]
1&b&a&-a&-a&a&-b&-1&\frac{b}{a}&-\frac{b}{a}&-\frac{b}{a}&\frac{b}{a}\\*[2mm]
1&a&c&a&-a&-a&-c&\frac{c}{a}&-1&\frac{c}{a}&-\frac{c}{a}&-\frac{c}{a}\\*[2mm]
1&-a&a&d&a&-a&-d&-\frac{d}{a}&\frac{d}{a}&-1&\frac{d}{a}&-\frac{d}{a}\\*[2mm]
1&-a&-a&a&e&a&-e&-\frac{e}{a}&-\frac{e}{a}&\frac{e}{a}&-1&\frac{e}{a}\\*[2mm]
1&a&-a&-a&a&f&-f&\frac{f}{a}&-\frac{f}{a}&-\frac{f}{a} &\frac{f}{a}&-1\\*[2mm]
1&-b&a&-a&-a&a&b&-1&-\frac{b}{a}&\frac{b}{a}&\frac{b}{a}&-\frac{b}{a}\\*[2mm]
1&a&-c&a&-a&-a&c&-\frac{c}{a}&-1&-\frac{c}{a}&\frac{c}{a}&\frac{c}{a}\\*[2mm]
1&-a&a&-d&a&-a&d&\frac{d}{a}&-\frac{d}{a}&-1&-\frac{d}{a}&\frac{d}{a}\\*[2mm]
1&-a&-a&a&-e&a&e&\frac{e}{a}&\frac{e}{a}&-\frac{e}{a}&-1&-\frac{e}{a}\\*[2mm]
1&a&-a&-a&a&-f&f&-\frac{f}{a}&\frac{f}{a}&\frac{f}{a} &-\frac{f}{a}&-1\\*[2mm]
1&-1&-1&-1&-1&-1&-1&1&1&1&1&1
\end{array}
\right]}\label{O12c}
\end{eqnarray}

By taking $a=b=c=d=e=f=1$ in the above three matrices we get
the corresponding  real  Hadamard matrices, that  have the form

\begin{eqnarray}
 H_{12a}=
{\small\left[\begin{array}{rrrrrrrrrrrr}
1&1&1&1&1&1&1&1&1&1&1&1\\
1&1&1&1&-1&-1&-1&-1&1&1&-1&-1\\
1&1&1&-1&1&-1&-1&1&-1&-1&1&-1\\
1&1&-1&1&-1&1&-1&1&-1&-1&-1&1\\
1&-1&1&-1&1&1&-1&-1&1&-1&-1&1\\
1&-1&-1&1&1&1&-1&-1&-1&1&1&-1\\
1&-1&1&1&-1&-1&1&-1&-1&-1&1&1\\
1&1&-1&-1&1&-1&1&-1&-1&1&-1&1\\
1&1&-1&-1&-1&1&1&-1&1&-1&1&-1\\
1&-1&1&-1&-1&1&1&1&-1&1&-1&-1\\
1&-1&-1&1&1&-1&1&1&1&-1&-1&-1\\
1&-1&-1&-1&-1&-1&-1&1&1&1&1&1
\end{array}
\right]}\label{H12a}
\end{eqnarray}

\begin{eqnarray}
 H_{12b}=
{\small\left[\begin{array}{rrrrrrrrrrrr}
1&1&1&1&1&1&1&1&1&1&1&1\\
1&1&1&1&-1&-1&-1&-1&1&1&-1&-1\\
1&1&1&-1&-1&1&-1&1&-1&-1&-1&1\\
1&1&-1&1&1&-1&-1&1&-1&-1&1&-1\\
1&-1&-1&1&1&1&-1&-1&-1&1&-1&1\\
1&-1&1&-1&1&-1&1&-1&1&-1&1&-1\\
1&-1&1&1&-1&-1&1&-1&-1&-1&1&1\\
1&1&-1&-1&-1&1&1&-1&-1&1&1&-1\\
1&1&-1&-1&1&-1&1&1&1&-1&-1&1\\
1&-1&-1&1&-1&1&1&1&1&-1&-1&1\\
1&-1&1&-1&1&-1&1&-1&1&-1&1&-1\\
1&-1&-1&-1&-1&-1&-1&1&1&1&1&1
\end{array}
\right]}\label{H12b}
\end{eqnarray}

\begin{eqnarray}
 H_{12c}=
{\small\left[\begin{array}{rrrrrrrrrrrr}
1&1&1&1&1&1&1&1&1&1&1&1\\
1&1&1&-1&-1&1&-1&-1&1&-1&-1&1\\
1&1&1&1&-1&-1&-1&1&-1&1&-1&-1\\
1&-1&1&1&1&-1&-1&-1&1&-1&1&-1\\
1&-1&-1&1&1&1&-1&-1&-1&1&-1&1\\
1&1&-1&-1&1&1&-1&1&-1&-1&1&-1\\
1&-1&1&-1&-1&1&1&-1&-1&1&1&-1\\
1&1&-1&1&-1&-1&1&-1&-1&-1&1&1\\
1&-1&1&-1&1&-1&1&1&-1&-1&-1&1\\
1&-1&-1&1&-1&1&1&1&1&-1&-1&-1\\
1&1&-1&-1&1&-1&1&-1&1&1&-1&-1\\
1&-1&-1&-1&-1&-1&-1&1&1&1&1&1
\end{array}
\right]}\label{H12c}
\end{eqnarray}

The complex Hadamard matrices written in  the standard form from \cite{TZ} are
\begin{eqnarray}
D_{12a}(a,b,c,d,e,f)=
 H_{12a}\circ  EXP\left( i\cdot R_{12}^{(6)}(a,b,c,d,e,f)\right)\label{h12a}
\end{eqnarray}

\begin{eqnarray}
D_{12b}(a,b,c,d,e,f)=
 H_{12b}\circ  EXP\left( i\cdot R_{12}^{(6)}(a,b,c,d,e,f)\right)\label{h12b}
\end{eqnarray}

\begin{eqnarray}
D_{12c}(a,b,c,d,e,f)=
 H_{12c}\circ  EXP\left( i\cdot R_{12}^{(6)}(a,b,c,d,e,f)\right)\label{h12c}
\end{eqnarray}

\begin{eqnarray}
R_{12}^{(6)}(a,b,c,d,e,f)=
\small{ \left[\begin{array}{cccccccccccc}
\bullet&\bullet&\bullet&\bullet&\bullet&\bullet&\bullet&\bullet&\bullet&\bullet
&\bullet&\bullet\\
\bullet&b&a&a&a&a&b&\bullet&b-a&b-a&b-a&b-a\\
\bullet&a&c&a&a&a&c&c-a&\bullet&c-a&c-a&c-a\\
\bullet&a&a&d&a&a&d&d-a&d-a&\bullet&d-a&d-a\\
\bullet& a&a&a&e&a&e&e-a&e-a&e-a&\bullet&e-a\\
\bullet&a&a&a&a&f&f&f-a&f-a&f-a&f-a&\bullet\\
\bullet&b&a&a&a&a&b&\bullet&b-a&b-a&b-a&b-a\\
\bullet&a&c&a&a&a&c&c-a&\bullet&c-a&c-a&c-a\\
\bullet&a&a&d&a&a&d&d-a&d-a&\bullet&d-a&d-a\\
\bullet&a&a&a&e&a&e&e-a&e-a&e-a&\bullet&e-a\\
\bullet&a&a&a&a&f&f&f-a&f-a&f-a&f-a&\bullet\\
\bullet&\bullet&\bullet&\bullet&\bullet&\bullet&\bullet&\bullet&\bullet&\bullet
&\bullet&\bullet
\end{array}
\right]}\label{R12}
\end{eqnarray}
where $(a,b,c,d,e,f) \in \R^6$.  The above relations have been put in their most symmetric form.

\section{Complex conference matrices}

From matrix (\ref{conf6}) for $p=q={\bf{i}}=\sqrt{-1}$ we get the conference matrix

\begin{eqnarray}
C_{6d}=\left[\begin{array}{rrrrrr}
0&1&1&1&1&1\\
1&0&{\bf i} &{\bf i}&-{\bf i} &-{\bf i} \\
1&{\bf i} &0&-{\bf i}&1&-1\\
1&{\bf i} &-{\bf i}&0&-1 &1 \\
1&-{\bf i}&-1 &1 &0&{\bf i}\\
1&-{\bf i} &1 &-1 &{\bf i}&0\\
\end{array}
\right]\label{conf6d}
\end{eqnarray}

The choice  $b= {\bf i},c=-{\bf i}$ in (\ref{conf6}) leads to the matrix

\begin{eqnarray}
C_{6e}=\left[\begin{array}{rrrrrr}
0&1&1&1&1&1\\
1&0&{\bf i} &{\bf i}&-{\bf i} &-{\bf i} \\
1&{\bf i} &0&-{\bf i}&-1&1\\
1&{\bf i} &-{\bf i}&0&1 &-1 \\
1&-{\bf i}&1 &-1 &0&{\bf i}\\
1&-{\bf i} &-1 &1 &{\bf i}&0\\
\end{array}
\right]\label{conf6e}
\end{eqnarray}

The  core of (\ref{conf6d}) and  (\ref{conf6e}) matrices, i.e. the $5\times 5$ inner matrix, is symetric. However it does exist  conference matrices whose core is skew-symmetric, namely the next two

\begin{eqnarray}
C_{6f}=\left[\begin{array}{rrrrrr}
0&1&1&1&1&1\\
1&0&1&{\bf i} &-{\bf i}&-1 \\
1&-1 &0&1&{\bf i} &-{\bf i}\\
1&-{\bf i} &-1&0&1 &{\bf i}\\
1&{\bf i}&-{\bf i} &-1 &0&1\\
1&1&{\bf i} &-{\bf i} &- 1&0\\
\end{array}
\right]\label{peter1}
\end{eqnarray}
and

\begin{eqnarray}
C_{6g}=\left[\begin{array}{rrrrrr}
0&1&1&1&1&1\\
1&0&1&-{\bf i} &{\bf i}&-1 \\
1&-1 &0&1&-{\bf i} &{\bf i}\\
1&{\bf i} &-1&0&1 &-{\bf i}\\
1&-{\bf i}&{\bf i} &-1 &0&1\\
1&1&-{\bf i} &{\bf i} &- 1&0\\
\end{array}
\right]\label{peter2}
\end{eqnarray}

The above four conference matrices lead to the following complex orthogonal matrices which depend on six non-zero complex parameters, as follows

\begin{eqnarray}
 O_{12d}(a,b,c,d,e,f)=
\small{\left[\begin{array}{crrrrrrrrrrr}
1&1&1&1&1&1&1&1&1&1&1&1\\*[2mm]
1&b&a&a&-a&-a&-b&-1&\frac{b}{a}&\frac{b}{a}&-\frac{b}{a}&-\frac{b}{a}\\*[2mm]
1&a&c&-a&-{\bf i\,}a&{\bf i\,}a&-c&\frac{c}{a}&-1&-\frac{c}{a}&-\frac{{\bf i\,}c}{a}&\frac{{\bf i\,}c}{a}\\*[2mm]
1&a&-a&d&{\bf i\,}a&{-\bf i\,}a&-d&\frac{d}{a}&-\frac{d}{a}&-1&\frac{{\bf i\,}d}{a}&-\frac{{\bf i\,}d}{a}\\*[2mm]
1&-a&{\bf i\,}a&-{\bf i\,}a&e&a&-e&-\frac{e}{a}&\frac{{\bf i\,}e}{a}&-\frac{{\bf i\,}e}{a}&-1&\frac{e}{a}\\*[2mm]
1&-a&-{\bf i\,}a&{\bf i\,}a&a&f&-f&-\frac{f}{a}&-\frac{{\bf i\,}f}{a}&\frac{{\bf i\,}f}{a} &\frac{f}{a}&-1\\*[2mm]
1&-b&a&a&-a&-a&b&-1&-\frac{b}{a}&-\frac{b}{a}&\frac{b}{a}&\frac{b}{a}\\*[2mm]
1&a&-c&-a&{-\bf i\,}a&{\bf i\,}a&c&-\frac{c}{a}&-1&\frac{c}{a}&\frac{{\bf i\,}c}{a}&-\frac{{\bf i\,}c}{a}\\*[2mm]
1&a&-a&-d&{\bf i\,}a&{\bf i\,}a&d&-\frac{d}{a}&\frac{d}{a}&-1&-\frac{{\bf i\,}d}{a}&\frac{{\bf i\,}d}{a}\\*[2mm]
1&-a&{\bf i\,}a&-{\bf i\,}a&-e&a&e&\frac{e}{a}&-\frac{{\bf i\,}e}{a}&\frac{{\bf i\,}e}{a}&-1&-\frac{e}{a}\\*[2mm]
1&-a&-{\bf i\,}a&{\bf i\,}a&a&-f&f&\frac{f}{a}&\frac{{\bf i\,}f}{a}&-\frac{{\bf i\,}f}{a} &-\frac{f}{a}&-1\\*[2mm]
1&-1&-1&-1&-1&-1&-1&1&1&1&1&1
\end{array}
\right]}\label{O12d}
\end{eqnarray}

\begin{eqnarray}
 O_{12e}(a,b,c,d,e,f)=
\small{\left[\begin{array}{crrrrrrrrrrr}
1&1&1&1&1&1&1&1&1&1&1&1\\*[2mm]
1&b&a&a&-a&-a&-b&-1&\frac{b}{a}&\frac{b}{a}&-\frac{b}{a}&-\frac{b}{a}\\*[2mm]
1&a&c&-a&{\bf i\,}a&-{\bf i\,}a&-c&\frac{c}{a}&-1&-\frac{c}{a}&\frac{{\bf i\,}c}{a}&-\frac{{\bf i\,}c}{a}\\*[2mm]
1&a&-a&d&-{\bf i\,}a&{\bf i\,}a&-d&\frac{d}{a}&-\frac{d}{a}&-1&-\frac{{\bf i\,}d}{a}&\frac{{\bf i\,}d}{a}\\*[2mm]
1&-a&-{\bf i\,}a&{\bf i\,}a&e&a&-e&-\frac{e}{a}&-\frac{{\bf i\,}e}{a}&\frac{{\bf i\,}e}{a}&-1&\frac{e}{a}\\*[2mm]
1&-a&{\bf i\,}a&-{\bf i\,}a&a&f&-f&-\frac{f}{a}&\frac{{\bf i\,}f}{a}&-\frac{{\bf i\,}f}{a} &\frac{f}{a}&-1\\*[2mm]
1&-b&a&a&-a&-a&b&-1&-\frac{b}{a}&-\frac{b}{a}&\frac{b}{a}&\frac{b}{a}\\*[2mm]
1&a&-c&-a&{\bf i\,}a&-{\bf i\,}a&c&-\frac{c}{a}&-1&\frac{c}{a}&-\frac{{\bf i\,}c}{a}&\frac{\bf i\,c}{a}\\*[2mm]
1&a&-a&-d&-{\bf i\,}{a}&{\bf i\,}a&d&-\frac{d}{a}&\frac{d}{a}&-1&\frac{{\bf i\,}d}{a}&-\frac{{\bf i\,}d}{a}\\*[2mm]
1&-a&-{\bf i\,}a&{\bf i\,}a&-e&a&e&\frac{e}{a}&\frac{{\bf i\,}e}{a}&-\frac{{\bf i\,}e}{a}&-1&-\frac{e}{a}\\*[2mm]
1&-a&{\bf i\,}a&-{\bf i\,}a&a&-f&f&\frac{f}{a}&-\frac{{\bf i\,}f}{a}&\frac{{\bf i\,}f}{a} &-\frac{f}{a}&-1\\*[2mm]
1&-1&-1&-1&-1&-1&-1&1&1&1&1&1
\end{array}
\right]}\label{O12e}
\end{eqnarray}

The structure of the following two matrices coming from $C_{6f}$ and  $C_{6d}$ is different from the structure of $O_{12d}$ and  $O_{12e}$
\begin{eqnarray}
 O_{12f}(a,b,c,d,e,f)=
\small{\left[\begin{array}{crrrrrrrrrrr}
1&1&1&1&1&1&1&1&1&1&1&1\\*[2mm]
1&b&-a&-{\bf i\,} a& {\bf i\,}a&a&-b&-1&\frac{b}{a}&-\frac{{\bf i\,}b}{a}&\frac{{\bf i\,}b}{a}&-\frac{b}{a}\\*[2mm]
1&a&c&-a&-{\bf i\,}a&{\bf i\,}a&-c&-\frac{c}{a}&-1&\frac{c}{a}&-\frac{{\bf i\,}c}{a}&\frac{{\bf i\,}c}{a}\\*[2mm]
1&{\bf i\,} a&a&d&-a&-{\bf i\,}a&-d&\frac{{\bf i\,}d}{a}&-\frac{d}{a}&-1&\frac{d}{a}&-\frac{{\bf i\,}d}{a}\\*[2mm]
1&-{\bf i\,}a&{\bf i\,}a&a&e&-a&-e&-\frac{{\bf i\,}e}{a}&\frac{{\bf i\,}e}{a}&-\frac{e}{a}&-1&\frac{e}{a}\\*[2mm]
1&-a&-{\bf i\,}a&{\bf i\,}a&a&f&-f&\frac{f}{a}&-\frac{{\bf i\,}f}{a}&\frac{{\bf i\,}f}{a} &-\frac{f}{a}&-1\\*[2mm]
1&-b&-a& -{\bf i\,}a&{\bf i\,}a&a&b&-1&-\frac{b}{a}&\frac{{\bf i\,}b}{a}&-\frac{{\bf i\,}b}{a}&\frac{b}{a}\\*[2mm]
1&a&-c&-a&-{\bf i\,}a&{\bf i\,}a&c&\frac{c}{a}&-1&-\frac{c}{a}&\frac{{\bf i\,}c}{a}&-\frac{\bf i\,c}{a}\\*[2mm]
1&{\bf i\,}a&a&-d&-a&-{\bf i\,}a&d&-\frac{{\bf i\,}d}{a}&\frac{d}{a}&-1&-\frac{d}{a}&\frac{{\bf i\,}d}{a}\\*[2mm]
1&-{\bf i\,}a&{\bf i\,}a&a&-e&-a&e&\frac{{\bf i\,}e}{a}&-\frac{{\bf i\,}e}{a}&\frac{e}{a}&-1&-\frac{e}{a}\\*[2mm]
1&-a&-{\bf i\,}a&{\bf i\,}a&a&-f&f&-\frac{f}{a}&\frac{{\bf i\,}f}{a}&-\frac{{\bf i\,}f}{a} &\frac{f}{a}&-1\\*[2mm]
1&-1&-1&-1&-1&-1&-1&1&1&1&1&1
\end{array}
\right]}\label{O12f}
\end{eqnarray}

\begin{eqnarray}
 O_{12g}(a,b,c,d,e,f)=
\small{\left[\begin{array}{crrrrrrrrrrr}
1&1&1&1&1&1&1&1&1&1&1&1\\*[2mm]
1&b&-a&{\bf i\,} a& -{\bf i\,}a&a&-b&-1&\frac{b}{a}&\frac{{\bf i\,}b}{a}&-\frac{{\bf i\,}b}{a}&-\frac{b}{a}\\*[2mm]
1&a&c&-a&{\bf i\,}a&-{\bf i\,}a&-c&-\frac{c}{a}&-1&\frac{c}{a}&\frac{{\bf i\,}c}{a}&-\frac{{\bf i\,}c}{a}\\*[2mm]
1&-{\bf i\,}a&a&d&-a&{\bf i\,}a&-d&-\frac{{\bf i\,}d}{a}&-\frac{d}{a}&-1&\frac{d}{a}&\frac{{\bf i\,}d}{a}\\*[2mm]
1&{\bf i\,}a&-{\bf i\,}a&a&e&-a&-e&\frac{{\bf i\,}e}{a}&-\frac{{\bf i\,}e}{a}&-\frac{e}{a}&-1&\frac{e}{a}\\*[2mm]
1&-a&{\bf i\,}a&-{\bf i\,}a&a&f&-f&\frac{f}{a}&\frac{{\bf i\,}f}{a}&-\frac{{\bf i\,}f}{a} &-\frac{f}{a}&-1\\*[2mm]
1&-b&-a& {\bf i\,}a&-{\bf i\,}a&a&b&-1&-\frac{b}{a}&-\frac{{\bf i\,}b}{a}&\frac{{\bf i\,}b}{a}&\frac{b}{a}\\*[2mm]
1&a&-c&-a&{\bf i\,}a&-{\bf i\,}a&c&\frac{c}{a}&-1&-\frac{c}{a}&-\frac{{\bf i\,}c}{a}&\frac{\bf i\,c}{a}\\*[2mm]
1&-{\bf i\,}a&a&-d&-a&{\bf i\,}a&d&\frac{{\bf i\,}d}{a}&\frac{d}{a}&-1&-\frac{d}{a}&-\frac{{\bf i\,}d}{a}\\*[2mm]
1&{\bf i\,}a&-{\bf i\,}a&a&-e&-a&e&-\frac{{\bf i\,}e}{a}&\frac{{\bf i\,}e}{a}&\frac{e}{a}&-1&-\frac{e}{a}\\*[2mm]
1&-a&{\bf i\,}a&-{\bf i\,}a&a&-f&f&-\frac{f}{a}&-\frac{{\bf i\,}f}{a}&\frac{{\bf i\,}f}{a} &\frac{f}{a}&-1\\*[2mm]
1&-1&-1&-1&-1&-1&-1&1&1&1&1&1
\end{array}
\right]}\label{O12g}
\end{eqnarray}

The corresponding Hadamard matrices obtained for $a=b=c=d=e=f=1$ are complex, and  have the following form

\begin{eqnarray}
 H_{12d}=
{\small\left[\begin{array}{crrrrrrrrrrr}
1&1&1&1&1&1&1&1&1&1&1&1\\
1&1&1&1&-1&-1&-1&-1&1&1&-1&-1\\
1&1&1&-1&-{\bf i}& {\bf i}&-1&1&-1&-1 &-{\bf i}& {\bf i}\\
1&1&-1&1&{\bf i}& -{\bf i}&-1&1&-1&-1 &{\bf i}&- {\bf i}\\
1&-1&{\bf i}&- {\bf i}&1&1&-1&-1&{\bf i}&- {\bf i}&-1&1\\
1&-1&-{\bf i}&{\bf i}&1&1&-1&-1&-{\bf i}& {\bf i}&-1&1\\
1&-1&1&1&-1&-1&1&-1&-1&-1&1&1\\
1&1&-1&-1&-{\bf i}& {\bf i}&-1&-1&-1&1 &{\bf i}& -{\bf i}\\
1&1&-1&-1&{\bf i}& -{\bf i}&-1&-1&-1&1 &-{\bf i}& {\bf i}\\
1&-1&{\bf i} &-{\bf i}&-1&1&1&1&{{\bf i}}&{\bf i}&-1&-1\\
1&-1&-{\bf i}&{\bf i}&1&-1&1&1&{\bf i}& -{\bf i}&-1&-1\\
1&-1&-1&-1&-1&-1&-1&1&1&1&1&1
\end{array}
\right]}\label{h12d}
\end{eqnarray}
\begin{eqnarray}
 H_{12e}=
{\small\left[\begin{array}{crrrrrrrrrrr}
1&1&1&1&1&1&1&1&1&1&1&1\\
1&1&1&1&-1&-1&-1&-1&1&1&-1&-1\\
1&1&1&-1&{\bf i}& -{\bf i}&-1&1&-1&-1 &{\bf i}& -{\bf i}\\
1&1&-1&1&-{\bf i}& {\bf i}&-1&1&-1&-1 &-{\bf i}& {\bf i}\\
1&-1&-{\bf i}& {\bf i}&1&1&-1&-1&-{\bf i}& {\bf i}&-1&1\\
1&-1&{\bf i}&-{\bf i}&1&1&-1&-1&{\bf i}& -{\bf i}&1&-1\\
1&-1&1&1&-1&-1&1&-1&-1&-1&1&1\\
1&1&-1&-1&{\bf i}& -{\bf i}&1&-1&-1&1 &-{\bf i}& {\bf i}\\
1&1&-1&-1&-{\bf i}& {\bf i}&1&-1&1&-1 &{\bf i}& -{\bf i}\\
1&-1&-{\bf i} & {\bf i}&-1&1&1&1&{{\bf i}}&-{\bf i}&-1&-1\\
1&-1&{\bf i}&-{\bf i}&1&-1&1&1&-{\bf i}& {\bf i}&-1&-1\\
1&-1&-1&-1&-1&-1&-1&1&1&1&1&1
\end{array}
\right]}\label{h12e}
\end{eqnarray}

\begin{eqnarray}
 H_{12f}=
{\small\left[\begin{array}{crrrrrrrrrrr}
1&1&1&1&1&1&1&1&1&1&1&1\\*[2mm]
1&1&-1&-{\bf i}&{\bf i} &1&-1&-1&1& -{\bf i}&{\bf i} &-1\\
1&1&1&-1&-{\bf i}& {\bf i}&-1&-1&-1&1 &-{\bf i}& {\bf i}\\
1&{\bf i} &1&1&-1&-{\bf i}&-1&{\bf i} &-1&-1&1&- {\bf i}\\
1&-{\bf i} &{\bf i}& 1&1&-1&-1&-{\bf i}&{\bf i}&-1&-1&1\\
1&-1&-{\bf i}&{\bf i}&1&1&-1&1&-{\bf i}& {\bf i}&-1&-1\\
1&-1&-1&-{\bf i} &{\bf i}&1&1&-1&-1&{\bf i}&-{\bf i}&1\\
1&1&-1&-1&-{\bf i}& {\bf i}&1&1&-1&-1&{\bf i}&- {\bf i}\\
1& {\bf i}&1&-1&-1& -{\bf i}&1&-{\bf i}&1&-1 &-1& {\bf i}\\
1&-{\bf i} & {\bf i}& 1&-1&-1&1&{\bf i}&{-\bf i} &1&-1&-1\\
1&-1&-{\bf i}&{\bf i}&1&-1&1&-1&{\bf i}& -{\bf i}&1&-1\\
1&-1&-1&-1&-1&-1&-1&1&1&1&1&1
\end{array}
\right]}\label{h12f}
\end{eqnarray}

\begin{eqnarray}
 H_{12g}=
{\small\left[\begin{array}{crrrrrrrrrrr}
1&1&1&1&1&1&1&1&1&1&1&1\\
1&1&-1&{\bf i}&-{\bf i} &1&-1&-1&1& {\bf i}&-{\bf i} &-1\\
1&1&1&-1&{\bf i}& -{\bf i}&-1&-1&-1&1 &{\bf i}& -{\bf i}\\
1&-{\bf i} &1&1&-1&{\bf i}&-1&-{\bf i} &-1&-1&1& {\bf i}\\
1&{\bf i} &-{\bf i}& 1&1&-1&-1&{\bf i}&-{\bf i}&-1&-1&1\\
1&-1&{\bf i}&-{\bf i}&1&1&-1&1&{\bf i}& -{\bf i}&-1&-1\\
1&-1&-1&{\bf i} &{-\bf i}&1&1&-1&-1&{-\bf i}&{\bf i}&1\\
1&1&-1&-1&{\bf i}& -{\bf i}&1&1&-1&-1&-{\bf i}&{\bf i}\\
1& -{\bf i}&1&-1&-1& {\bf i}&1&{\bf i}&1&-1 &-1& -{\bf i}\\
1&{\bf i} & {-\bf i}& 1&-1&-1&1&{-\bf i}&{\bf i} &1&-1&-1\\
1&-1&{\bf i}&-{\bf i}&1&-1&1&-1&-{\bf i}& {\bf i}&1&-1\\
1&-1&-1&-1&-1&-1&-1&1&1&1&1&1
\end{array}
\right]}\label{h12g}
\end{eqnarray}
Similarly to the preceding case the complex Hadamard matrices depending on six phases can be written as

\begin{eqnarray}
D_{12d}(a,b,c,d,e,f)=
 H_{12d}\circ  EXP\left( i\cdot R_{12}^{(6)}(a,b,c,d,e,f)\right)\label{d12d}
\end{eqnarray}
\begin{eqnarray}
D_{12e}(a,b,c,d,e,f)=
 H_{12e}\circ  EXP\left( i\cdot R_{12}^{(6)}(a,b,c,d,e,f)\right)\label{d12e}
\end{eqnarray}
\begin{eqnarray}
D_{12f}(a,b,c,d,e,f)=
 H_{12f}\circ  EXP\left( i\cdot R_{12}^{(6)}(a,b,c,d,e,f)\right)\label{d12f}
\end{eqnarray}
\begin{eqnarray}
D_{12g}(a,b,c,d,e,f)=
 H_{12g}\circ  EXP\left( i\cdot R_{12}^{(6)}(a,b,c,d,e,f)\right)\label{d12g}
\end{eqnarray}
with $(a,b,c,d,e,f) \in \R^6$.

The above four orthogonal matrices $O_{12d}-O_{12g}$ have a special dependence on the imaginary unit, ${\bf i}$, i.e. $ a,\, {\bf i}\,a, b,\, {\bf i}\,b $, etc,  such that in these cases  does not exist  reparametrizations  to bring them to the  similar form of  $O_{12a}-O_{12c}$ matrices.

In the following we make use of the conference matrix (\ref{conf6}) that has a non linear dependence on parameters, and this propertty implies a new complex parameter in the corresponding Hadamard matrix.
Indeed the matrix (\ref{conf6}) provides the following matrix which depend on seven parameters.

\begin{eqnarray}
 O_{12h}(a,b,c,d,e,f,g)=
{\small\left[\begin{array}{crrrrrrrrrrr}
1&1&1&1&1&1&1&1&1&1&1&1\\*[2mm]
1&b&a& a&-a&-a&-b&-1&\frac{b}{a}&\frac{b}{a}&-\frac{b}{a}&-\frac{b}{a}\\*[2mm]
1&a&c&-a&\frac{a}{g}&-\frac{a}{g}&-c&\frac{c}{a}&-1&-\frac{c}{a}&\frac{c}{ag}&-\frac{c}{ag}\\*[2mm]
1&a&-a&d&-\frac{a}{g}&\frac{a}{g}&-d&\frac{d}{a}&-\frac{d}{a}&-1&-\frac{d}{ag}&\frac{d}{ag}\\*[2mm]
1&-a&ag&-ag &e&a&-e&-\frac{e}{a}&\frac{eg}{a}&-\frac{eg}{a}&-1&\frac{e}{a}\\*[2mm]
1&-a&-ag&ag&a&f&-f&-\frac{f}{a}&-\frac{fg}{a}&\frac{fg}{a} &\frac{f}{a}&-1\\*[2mm]
1&-b&a&a&-a&-a&b&-1&-\frac{b}{a}&-\frac{b}{a}&\frac{b}{a}&\frac{b}{a}\\*[2mm]
1&a&-c&-a&\frac{a}{g}&-\frac{a}{g} &c&-\frac{c}{a}&-1&\frac{c}{a}&-\frac{c}{ag}&\frac{c}{ag}\\*[2mm]
1&a&-a&-d&-\frac{a}{g}&\frac{a}{g} &d&-\frac{d}{a}&\frac{d}{a}&-1&\frac{d}{ag}&-\frac{d}{ag}\\*[2mm]
1&-a&ag&- ag &-e&a&e&\frac{e}{a}&-\frac{e}{ag}&\frac{e}{ag}&-1&-\frac{e}{a}\\*[2mm]
1&-a&-ag&ag &a&-f&f&\frac{f}{a}&\frac{fg}{a}&-\frac{fg}{a} &-\frac{f}{a}&-1\\*[2mm]
1&-1&-1&-1&-1&-1&-1&1&1&1&1&1
\end{array}
\right]}\label{O12h}
\end{eqnarray}

\begin{eqnarray}
R_{12}^{(7)}(a,b,c,d,e,f,g)=~~~~~~~~~~~~~~~~~~~~~~~~~~~~~~~~~~~~~~~~~~~~~~~~~~~~~~~~~~~~~~~~~~~~~~~~~~~~~~~~~~~~~~~~~\nonumber\\
\small{\left[\begin{array}{cccccccccccc}
\bullet&\bullet&\bullet&\bullet&\bullet&\bullet&\bullet&\bullet&\bullet&\bullet
&\bullet&\bullet\\
\bullet&b&a&a&a&a&b&\bullet&b-a&b-a&b-a&b-a\\
\bullet&a&c&a&a-g&a-g&c&c-a&\bullet&c-a&c-a-g&c-a-g\\
\bullet&a&a&d&a-g&a-g&d&d-a&d-a&\bullet&d-a-g&d-a-g\\
\bullet& a&a+g&a+g&e&a&e&e-a&e+g-a&e+g-a&\bullet&e-a\\
\bullet&a&a+g&a+g&a&f&f&f-a&f+g-a&f+g-a&f-a&\bullet\\
\bullet&b&a&a&a&a&b&\bullet&b-a&b-a&b-a&b-a\\
\bullet&a&c&a&a-g&a-g&c&c-a&\bullet&c-a&c-a-g&c-a-g\\
\bullet&a&a&d&a-g&a-g&d&d-a&d-a&\bullet&d-a-g&d-a-g\\
\bullet&a&a+g&a+g&e&a&e&e-a&e-a-g&e-a-g&\bullet&e-a\\
\bullet&a&a+g&a+g&a&f&f&f-a&f+g-a&f+g-a&f-a&\bullet\\
\bullet&\bullet&\bullet&\bullet&\bullet&\bullet&\bullet&\bullet&\bullet&\bullet
&\bullet&\bullet
\end{array}
\right]}\label{R127}
\end{eqnarray}
where $(a,b,c,d,e,f,g) \in \R^7$. The standard form of the  matrix (\ref{O12h}) is

\begin{eqnarray}
D_{12h}(a,b,c,d,e,f,g)=
 H_{12a}\circ  EXP\left( i\cdot R_{12}^{(7)}(a,b,c,d,e,f,g)\right)\label{H12ah}
\end{eqnarray}

All the 12-dimensional complex matrices found in the paper have a beautiful and rigid symmetry that allows to easily check their (non)equivalence.

\section{Conclusion}
In this paper we provided a procedure to find parametrizations of complex inverse orthogonal matrices by doubling the size of complex inverse orthogonal conference matrices. Our construction shows that all the existing $n$-dimensional real conference matrices lead to continuos families of orthogonal matrices depending on $n$ complex parameters. These matrices generate a large family of real Hadamard matrices by making $a_i=\pm 1$, $i=1,\,\dots,n$, and in this family we have to look for nonequivalent real Hadamard matrices. Thus it gets possible to find at least a minimal set of nonequivalent real Hadamard matrices for any $2n \ge 16$.

Our construction for $n = 6$ showed that by using complex conference matrices the number of nonequivalent complex Hadamard matrices increased significantly. Thus a main problem is to find new methods for  construction of  complex conference matrices for  higher dimensions. In our paper we used an ad-hoc approach starting with conference matrices depending on two independent phases, and transforming them into Sylvester conference matrices.



\end{document}